**Raigul Zheldibayeva**

**Zhetysu University named after I. Zhansugurov; University of Illinois Urbana-Champaign - Bolashak International Scholarship, Kazakhstan**

# THE IMPACT OF AI AND PEER FEEDBACK ON RESEARCH WRITING SKILLS: A STUDY USING THE CGSCHOLAR PLATFORM AMONG KAZAKHSTANI SCHOLARS

**Abstract:** This research studies the impact of AI and Peer feedback on the academic writing development of Kazakhstani scholars using the CGScholar platform − the product of cutting-edge research and development into collaborative learning, big data, and artificial intelligence developed by educators and computer scientists at the University of Illinois Urbana-Champaign (UIUC). The study aimed to find out how familiarity with AI tools and peer feedback processes impacts participants' openness to incorporating feedback into their academic writing. The study involved 36 scholars enrolled in a scientific internship focused on education at the University of UIUC. A survey with 15 questions with multiple-choice, Likert scale, and open-ended questions was employed to collect a data. The survey was conducted via Google Forms in both English and Russian to ensure linguistic accessibility. Demographic information such as age, gender, and first language were collected to provide a detailed understanding of the data. The analysis revealed a moderate positive correlation between familiarity with AI tools and openness to making changes based on feedback, and a strong positive correlation between research writing experience and expectations of peer feedback, especially in the area of research methodology. These results show that participants are open-minded to AI-assisted feedback; however, they still highly appreciate peer input, especially regarding methodological guidance. This study demonstrates the potential benefits of integrating AI tools with traditional feedback mechanisms to improve research writing quality in academic settings. Further research is recommended to evaluate the long-term impact of AI and peer feedback on academic writing skills, particularly through longitudinal studies that assess skill retention over multiple feedback cycles. Additionally, expanding the study to include a more diverse academic audience will provide deeper insights into how feedback mechanisms function across different research cultures and disciplines.

**Introduction**

In today's digital era, technology plays an increasingly significant role in shaping how we interact with the world and each other. These interactions are often facilitated by networked computing, which enhances human intelligence and allows for greater efficiency despite the foundational simplicity of binary computing systems (Cope & Kalanztis, 2019). One of the key advancements made possible by this technology is the use of artificial intelligence (AI) in educational settings (Kong et al.,2022). AI tools are now commonly used to enhance academic practices such as feedback and assessments, which are critical to improving students' learning experiences.

Despite its efficiency in correcting linguistic errors and improving structural coherence, AI-generated feedback, struggles in evaluating critical thinking, argument development, and ethical considerations in academic writing. Some studies (Abel et al.,2022) indicate that while AI can process big amounts of textual data and provide instant feedback, it lacks the ability to assess the reasoning and depth of argumentation that are essential for academic writing. Thus, according to Varshney et al. (2019), AI's reliance on pre-trained algorithms means it does not engage in reasoning, contextual understanding, or counterargument evaluation. Additionally, ethical concerns arise in AI-assisted writing due to algorithmic bias, potential plagiarism misidentifications, and the risk of over-reliance on machine-generated text (Illia et al.,2022). Thus, while AI models are succeeding in detecting surface-level errors, they struggle with specific fields and fail to accommodate the detailed argumentation required in different academic fields. Consequently, a hybrid model integrating AI for linguistic accuracy and peer- feedback for higher-order thinking assessment could be essential for balanced, meaningful, and ethically appropriate academic evaluation system.

CGScholar, an educational platform integrating AI and learning analytics, exemplifies how new technologies are transforming learning environments. Initially developed as part of an Australian government project focused on social literacy, CGScholar has since evolved into a global platform that encourages collaboration, idea-sharing, and interdisciplinary dialogue (Cope & Kalanztis, 2019). Managed by Common Ground Research Networks, the platform supports over 350,000 accounts, with more than 45,000 monthly users by 2023. CGScholar is used in a wide range of educational settings, from K-12 literacy programs to higher education courses in fields such as engineering and medicine [6, p.7]. One of its key features is its ability to provide AI-driven and peer feedback, which aims to enhance research writing and academic quality. Both AI and peer feedback could be found in learning metrics which is represented in Figure 1.

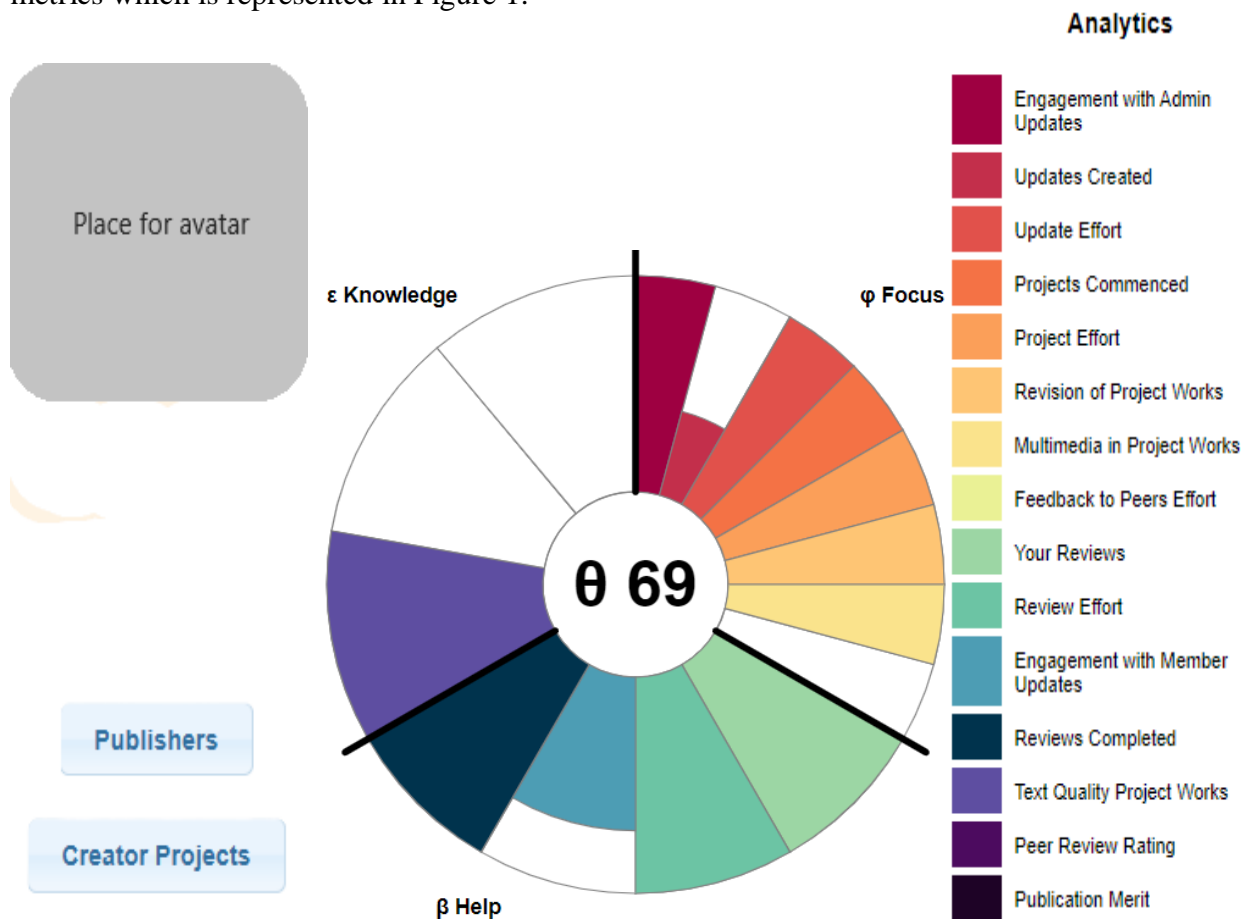

Figure 1. The visualization of learning metrics. https://cgscholar.com/identity/dashboard

Feedback has always been an essential component of academic success. Traditionally, feedback and assessments were delivered through written exams, multiple-choice tests, or short-answer questions (Santo, 2011). While these methods are effective to some extent, they often focus on memorization rather than critical thinking. Moreover, assessments such as multiple-choice or true/false questions can fail to fully evaluate higher-level cognitive skills [7, p.3]. Written responses, on the other hand, are more insightful but are often time-consuming to grade and subject to variability in interpretation [6, p.9].

Digital platforms like CGScholar address these limitations by leveraging AI for automated feedback and scoring, allowing for more efficient and immediate evaluation (Cope & Kalanztis, 2017). CGScholar not only facilitates AI-driven feedback but also encourages peer reviews and self-assessments, fostering a collaborative learning environment (Montebello et al., 2019). Furthermore, the platform is built on the principles of design-based research, which follows a process involving analysis, design, evaluation, and refinement to ensure continuous improvement (Allagui, 2023). This methodology helps balance theoretical ideals and practical applications through ongoing feedback loops.

Given the increasing integration of technology into educational settings, understanding how AI and peer feedback contribute to research quality is crucial. This study investigates the role of AI and peer feedback mechanisms on the CGScholar platform, specifically focusing on Kazakhstani scholars. By examining their expectations and experiences, the research aims to evaluate the strengths and limitations of these feedback mechanisms in enhancing research writing. The findings will provide insights into the growing role of AI in academic settings and its potential to transform traditional learning processes. Research writing skills, in the context of this study, refer to the ability of scholars to develop coherent, structured, and academically sound written works that meet the criteria outlined in the CGScholar platform's rubric (Cope & Kalanztis, 2023). These skills encompass clarity, argumentation, evidence integration, and adherence to academic standards. The study utilizes the Knowledge Processes Rubric to evaluate these components systematically.

**Research methodology**

*Research Design.* This research employed a design-based research (DBR) approach, focusing on the iterative development and evaluation of feedback mechanisms, specifically AI and peer feedback on the CGScholar platform. The study involved scholars from Kazakhstan's Pedagogy cohort, who were enrolled in the course "Transformations in Higher Education: Teaching, Learning, and Research in a Time of Disruptive Change". This course examined the impact of technological changes in higher education and explored innovative strategies for adapting to disruptive teaching, learning, and research trends. The research was conducted over one academic semester, with participants engaged in in-person and online sessions designed to examine critical themes in higher education, such as e-learning, artificial intelligence, and learner diversity. The scholars participated in multiple project updates, allowing for structured feedback from AI tools and peers on their research projects.

*Participants.* The participants in this study were 36 Kazakhstani scholars from the Pedagogy cohort, completing a scientific internship at the UIUC. These scholars participated in the course "Transformations in Higher Education: Teaching, Learning, and Research in a Time of Disruptive Change," which provided a structured environment for engaging with both AI and peer feedback on their research projects. To ensure scholars can effectively use the CGScholar platform and its tools, a comprehensive video tutorial (Zheldibayeva, 2024) has been prepared in Kazakh to help participants quickly sign up and explore the platform. The tutorial demonstrates the registration process, explains key features, and guides users through the platform's functionality, ensuring they can access and engage with the environment seamlessly.

*Course Structure*

The course was designed around weekly sessions, alternating between in-person and online formats. The key sessions related to this research include:

1. Project Updates: Scholars provided regular updates on their project progress (Weeks 3, 6, 7, and 11). Each update involved feedback loops with AI and peer reviewers to track improvements and ensure iterative development.
2. AI Feedback: During week 7, scholars received feedback from AI tools on their research drafts. This phase was crucial for evaluating the role of AI in identifying structural, grammatical, and coherence-related issues in the writing process.
3. Peer Feedback: During week 9, scholars engaged in a structured peer review process where they provided and received feedback on each other's work. This step was designed to foster collaborative learning and refine scholarly work through community-driven insights.
4. Final drafts: During the last week 12, course participants are expected to submit their final drafts, which have been revised based on AI and peer feedback. These drafts then would be reflected upon in the final session of the «Reflections on Peer and AI Collaborations» course.

Throughout the course, scholars received AI-generated feedback through the CGScholar platform, which utilized a detailed rubric to evaluate various aspects of academic writing. This rubric categorized feedback into six key areas:

1) Experience: This category evaluates how well participants of the course connect their personal or professional experiences to the topic. AI feedback assesses the clarity and effectiveness of linking the scholar's background, motivation, and experiences to the subject matter.

2) Empirical Evidence: AI feedback evaluates the presence and quality of empirical data or foundational information. It examines how effectively the scholar supports their arguments with research-based evidence.

3) Conceptualization: This feedback focuses on using fundamental concepts and the clarity of their definitions. AI also suggests ways to expand or refine the range of concepts used in the work to improve its depth and coherence.

4) Analysis: This section of the rubric addresses the reasoning, logic, and critical analysis presented by the scholar. AI evaluates whether the scholar's explanations are clear and demonstrate awareness of potential critiques or alternative theories.

5) Application: AI feedback in this area assesses how well the scholar translates theoretical ideas into practical applications.

6) Presentation: It evaluates the clarity of the presentation, as it is important that sources are cited correctly and that the scholar's voice stands out clearly from their references. All these six categories are presented in Figure 2.

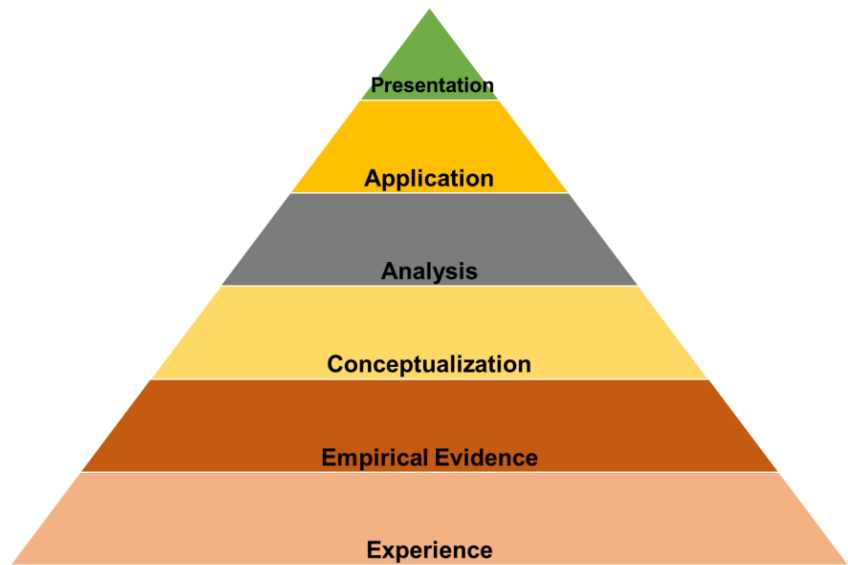

Figure 2. Rubrics categorization

The rubric was used to make both AI and peer feedback more organized, structured and corresponding to academic standards.

Data Collection. Our survey included 15 questions aiming to collect both quantitative and qualitative data. Multiple-choice questions, Likert-scale questions as well as open-ended questions were employed to achieve appropriate results. While multiple-choice and Likert-scale questions focused on participants' familiarity with AI tools, open-ended questions were devoted to peer feedback processes, openness to feedback, and experience with writing research papers. The questions were designed carefully in order to get adequate, measurable answers, which could provide information about participants' attitudes and future expectations related to both AI and peer feedback. The survey was conducted online in two languages, Russian and English, through Google Forms platform. This bilingual approach was used to minimize language barriers, as it was very important to ensure that they perceived all the questions and enabled participants to express their thoughts freely, improving the reliability of the data collected. In addition, demographic information such as age, gender, citizenship, and first language was collected. The collected data allowed us to analyze in more detail how these factors might influence participants' familiarity with digital tools and their openness to feedback. A sample of the results collected through Google Forms is presented in Figure 3.

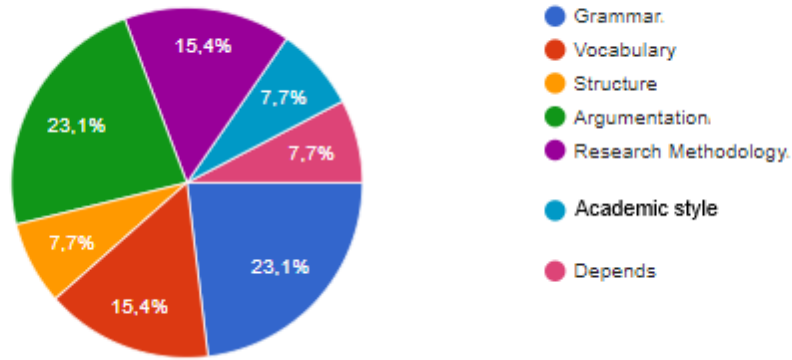

Figure 3. The fragment from Google Forms survey results.

Data for this study was gathered through a semi-structured preliminary survey conducted with all participants before they engaged with the feedback mechanisms. The survey focused on collecting demographic information, participants' prior experience with research writing, and their expectations regarding AI and peer feedback. Key focus areas included confidence in writing research papers in English, familiarity with AI tools and peer feedback mechanisms, challenges in academic writing and expectations about how AI and peer feedback might impact their research.

*Data Analysis.* Descriptive statistics analyzed quantitative data from the preliminary survey to provide an overview of participants' basic knowledge and expectations. Qualitative data was used for thematic analysis to identify key themes related to the perceived value of AI and peer feedback in academic writing.

*Limitations.* The given research provides valuable insights into the impact of AI and peer feedback on research writing skills. However, several limitations must be underlined: the study primarily examined the pre-survey feedback phase without conducting follow-up evaluations. This limits the ability to assess long-term improvements in writing skills or determine whether the initial feedback translated into sustained development over time. With data, measuring how effectively participants applied the input or how their skills evolved in subsequent writing tasks is challenging sample size: The relatively small sample of Kazakhstani scholars may limit the validation of the findings to other academic settings or demographic groups. Because of the limited diversity of the participant pool, individual responses could significantly influence the results.

Bilingual survey design: Surveying both English and Russian introduced potential differences in understanding and interpretation, which may have affected the accuracy of survey responses.

Also we highly recommend future research which could address these gaps by adopting a longitudinal approach that includes follow-up assessments to track participants' progress. This could involve evaluating changes in writing quality after multiple rounds of AI- assisted and peer feedback to better understand how these mechanisms impact skill development. Broadening the sample size to include a more diverse group of participants and ensuring consistency in language options would also enhance the reliability of the results academic context.

**Results**

*Demographic Characteristics.* The survey was conducted among 36 participants, all of whom were female citizens of Kazakhstan. The majority (58%) of participants were aged 46 and above, with 25% aged 26-35 and 17% aged 36-45. All participants identified Kazakh as their first language. Demographic characteristics are structured in Table 1.

| Characteristic | n | Percentage |
| --- | --- | --- |
| Age 46 and above | 21 | 58 |
| Age 36-45 | 6 | 17 |
| Age 26-35 | 9 | 25 |
| Female | 36 | 100 |
| Kazakhstan | 36 | 100 |
| Kazakh (first language) | 36 | 100 |

Table 1. Demographic characteristics

*Familiarity with AI and peer feedback.* The given data shows that the focus group generally has a low level of familiarity with AI tools, with an average score of 1.92 based on a scale from 1 (not familiar at all) to 5 (very familiar). This low awareness suggests that although AI tools are becoming increasingly available in academic and professional settings, they may not yet be widely accepted or understood by participants. The reasons may be limited access to these technologies, insufficient familiarity with AI tools in their academic programs, or lack of confidence in their effective use reason may be limited access to these technologies, insufficient familiarity with AI tools in their academic programs, or lack of confidence in the effective use of them in their research practice. In contrast, the average awareness score for peer feedback processes showed moderate awareness, indicating 3.67. These results show that participants are more exposed to traditional peer feedback than to AI support feedback. Peer feedback is probably a more reliable practice in their educational environment, which makes it more accessible and familiar. Familiarity with AI and peer feedback highlights not only the gap but also a potential area for growth, as integrating AI tools could complement and enhance traditional peer feedback methods.

*Openness to feedback.* Participants scored an average of 3.83 for the category openness to feedback, indicating that they are generally receptive to incorporating feedback into their research work. This level of openness is inspiring because it reflects an enthusiasm to improve and adapt based on feedback from various sources. We can see its particular significance as openness to feedback is a critical factor in successful learning and skill development, especially in academic writing. It suggests that participants might be receptive to AI-generated feedback once they become more familiar with the tools, given their positive attitude toward feedback.

*Experience in research writing.* Participants reported an average score of 3.5 for expertise in research writing, reflecting a moderate level of experience. This record clearly indicates that while the focus group shows some skills in academic writing in English, and it is a good achievement, considering that it is a foreign language for them, there is still a need for further improvements. A moderate level of experience suggests that participants may be in a transitional stage, a stage where they gain confidence and mastery in academic writing. By increasing their openness to feedback, this group could benefit from targeted interventions such as AI-enabled writing tools (Cope et al., 2023; Anderson, 2023) or peer feedback workshops to further improve their skills. The following results of the correlation analysis can be found in Table 2.

| Category | Value |
| --- | --- |
| AI Familiarity & Peer Feedback | 0.5 |
| Writing Experience & Peer Feedback | 0.7 |
| AI Familiarity & Openness | 1.0 |
| Experience in Writing | 3.0 |
| Openness to Feedback | 3.5 |
| Familiarity with Peer Feedback | 3.5 |
| Familiarity with AI | 2.5 |

Table 2. Research results: average scores and correlations

- A moderate positive correlation (r = 0.35) was found between familiarity with AI tools and openness to making changes based on feedback.
- A strong positive correlation (r = 0.68) was observed between experience in research writing and expectations for peer feedback, particularly in research methodology.
- A very strong correlation (r = 0.82) was also found between familiarity with AI tools and expectations for peer feedback on research methodology.

**Discussion**

This research study examined the relationship between participants' familiarity with AI tools, peer feedback, and openness to making changes based on feedback and incorporating them in research writing. The results provided some valuable insights into how these factors interact within the academic environment, particularly in the context of Kazakhstani scholars.

Key Findings. One of the significant finding we can underline could be a moderate positive correlation (0.35) between familiarity with AI tools and openness to make changes based on received feedback. This could point to the fact that participants who are more familiar with AI tools tend to be more receptive to feedback. However, the correlation is not particularly strong, suggesting that factors other than familiarity with AI tools may play a larger role in determining openness to feedback. This highlights a potential area for further research, particularly in understanding the role of technology in shaping attitudes toward feedback (Kalantzis & Cope, 2015). Similarly, a strong positive correlation (0.68) was observed between experience in research writing and expectations for peer feedback. This suggests that participants with more writing experience tend to have higher expectations for peer feedback, especially when it comes to receiving constructive advice on research methodology. This aligns with previous studies (Kalantzis & Cope, 2015; Fitria, 2021), which indicate that more experienced researchers often seek peer input to refine their methodologies and improve the quality of their work. Finally, the very strong correlation (0.82) between familiarity with AI tools and expectations for peer feedback on research methodology further supports the idea that those who are comfortable with AI tools also recognize the value of high-quality peer feedback in academic work.

These findings emphasize the interconnected nature of experience, familiarity, and openness, offering valuable insights (Alagui, 2023; Saini, 203; Castro, 2023) for improving academic practices.

Comparison with Previous Research. The results align with previous studies that suggest AI tools can improve structural elements of academic writing, such as grammar and clarity, but are less effective in addressing deeper aspects like argument coherence and research relevance. The finding that participants with more experience in research writing value peer feedback more highly is consistent with the literature that emphasizes the role of collaborative learning in improving academic outputs.

*Empirical Evidence.* The research effectively integrates empirical evidence to highlight the significance of AI and peer feedback in academic settings. The survey conducted among Kazakhstani scholars revealed a moderate positive correlation between familiarity with AI tools and openness to feedback. This finding provides a quantitative measure supporting the essay's argument that familiarity with technology can influence scholars' receptiveness to feedback mechanisms. In this research, we combined both qualitative and quantitative approaches into the research design, which could provide a strong empirical background, giving a detailed understanding of participants' perceptions and experiences.

However, it should be noted that since this study focused only on Kazakhstani scholars, the empirical scope of this study is currently limited. Future research could examine the way these findings apply to a broader academic audience. For example, investigating diverse samples across different disciplines and cultural contexts will provide a more detailed perspective on both AI and peer feedback's role in academic writing. Moreover, since this study did not measure long-term skill retention, future research could implement longitudinal assessments to track the sustained impact of AI-generated feedback and peer feedback on writing development over time. Future studies are expected to include larger and more diverse samples to achieve a deeper understanding of how feedback mechanisms can improve academic writing within different disciplines and cultural contexts.

**Conclusion**

This study explored the impact of AI-generated and peer feedback on the research writing skills of Kazakhstani scholars using the CGScholar platform. Through a combination of descriptive and correlation analyses, we examined how familiarity with AI tools, experience in academic writing, and expectations for peer feedback influence scholars' openness to incorporating feedback into their work.

The findings reveal that although participants generally had limited familiarity with AI tools, they demonstrated a moderate level of openness to feedback. More experienced scholars tended to rely more on peer feedback, particularly for improving research methodology. The correlation analysis further confirmed these trends. A moderate positive correlation was found between familiarity with AI tools and openness to making changes based on feedback. A strong correlation was observed between research writing experience and peer feedback expectations, particularly in research methodology. Additionally, a very strong correlation was found between familiarity with AI tools and expectations for peer feedback, suggesting that those who are comfortable with AI tools also recognize the value of peer feedback in academic work.

These results show some benefits of integrating AI-generated feedback into academic settings. However, since the study focused only on Kazakhstani scholars, future research should explore its applicability to a broader academic audience. Moreover, since long-term skill retention was not measured, future studies could include longitudinal assessments to track the sustained impact of AI and peer feedback on writing development. Expanding the sample size and diversity will provide deeper insights into how feedback mechanisms improve academic writing across different disciplines and cultural contexts.